\documentclass[twocolumn,prl,superscriptaddress,showpacs]{revtex4}
\usepackage{graphicx}
\begin{document}
\title{Exclusive $\rho^0$ meson electroproduction from hydrogen
at CLAS}

\newcommand*{\ORSAY }{ Institut de Physique Nucleaire ORSAY, Orsay, France} 
\affiliation{\ORSAY } 

\newcommand*{\INFNFR }{ INFN, Laboratori Nazionali di Frascati, Frascati, Italy} 
\affiliation{\INFNFR } 

\newcommand*{\SACLAY }{ CEA-Saclay, Service de Physique Nucl\'eaire, F91191 Gif-sur-Yvette,Cedex, France} 
\affiliation{\SACLAY } 

\newcommand*{\JLAB }{ Thomas Jefferson National Accelerator Facility, Newport News, Virginia 23606} 
\affiliation{\JLAB } 

\newcommand*{\WM }{ College of William and Mary, Williamsburg, Virginia 23187-8795} 
\affiliation{\WM } 

\newcommand*{\ASU }{ Arizona State University, Tempe, Arizona 85287-1504} 
\affiliation{\ASU } 

\newcommand*{\UCLA }{ University of California at Los Angeles, Los Angeles, California  90095-1547} 
\affiliation{\UCLA } 

\newcommand*{\CMU }{ Carnegie Mellon University, Pittsburgh, Pennsylvania 15213} 
\affiliation{\CMU } 

\newcommand*{\CUA }{ Catholic University of America, Washington, D.C. 20064} 
\affiliation{\CUA } 

\newcommand*{\CNU }{ Christopher Newport University, Newport News, Virginia 23606} 
\affiliation{\CNU } 

\newcommand*{\UCONN }{ University of Connecticut, Storrs, Connecticut 06269} 
\affiliation{\UCONN } 

\newcommand*{\DUKE }{ Duke University, Durham, North Carolina 27708-0305} 
\affiliation{\DUKE } 

\newcommand*{\ECOSSEE }{ Edinburgh University, Edinburgh EH9 3JZ, United Kingdom} 
\affiliation{\ECOSSEE } 

\newcommand*{\FIU }{ Florida International University, Miami, Florida 33199} 
\affiliation{\FIU } 

\newcommand*{\FSU }{ Florida State University, Tallahassee, Florida 32306} 
\affiliation{\FSU } 

\newcommand*{\GEISSEN }{ Physikalisches Institut der Universitaet Giessen, 35392 Giessen, Germany} 
\affiliation{\GEISSEN } 

\newcommand*{\GWU }{ The George Washington University, Washington, DC 20052} 
\affiliation{\GWU } 

\newcommand*{\ECOSSEG }{ University of Glasgow, Glasgow G12 8QQ, United Kingdom} 
\affiliation{\ECOSSEG } 

\newcommand*{\INFNGE }{ INFN, Sezione di Genova, 16146 Genova, Italy} 
\affiliation{\INFNGE } 

\newcommand*{\ITEP }{ Institute of Theoretical and Experimental Physics, Moscow, 117259, Russia} 
\affiliation{\ITEP } 

\newcommand*{\JMU }{ James Madison University, Harrisonburg, Virginia 22807} 
\affiliation{\JMU } 

\newcommand*{\KYUNGPOOK }{ Kyungpook National University, Daegu 702-701, South Korea} 
\affiliation{\KYUNGPOOK } 

\newcommand*{\MIT }{ Massachusetts Institute of Technology, Cambridge, Massachusetts  02139-4307} 
\affiliation{\MIT } 

\newcommand*{\UMASS }{ University of Massachusetts, Amherst, Massachusetts  01003} 
\affiliation{\UMASS } 

\newcommand*{\UNH }{ University of New Hampshire, Durham, New Hampshire 03824-3568} 
\affiliation{\UNH } 

\newcommand*{\NSU }{ Norfolk State University, Norfolk, Virginia 23504} 
\affiliation{\NSU } 

\newcommand*{\OHIOU }{ Ohio University, Athens, Ohio  45701} 
\affiliation{\OHIOU } 

\newcommand*{\ODU }{ Old Dominion University, Norfolk, Virginia 23529} 
\affiliation{\ODU } 

\newcommand*{\PENN }{ Penn State University, University Park, Pennsylvania 15260} 
\affiliation{\PENN } 

\newcommand*{\PITT }{ University of Pittsburgh, Pittsburgh, Pennsylvania 15260} 
\affiliation{\PITT } 

\newcommand*{\RPI }{ Rensselaer Polytechnic Institute, Troy, New York 12180-3590} 
\affiliation{\RPI } 

\newcommand*{\ROMA }{ Universita' di ROMA III, 00146 Roma, Italy} 
\affiliation{\ROMA } 

\newcommand*{\RICE }{ Rice University, Houston, Texas 77005-1892} 
\affiliation{\RICE } 

\newcommand*{\URICH }{ University of Richmond, Richmond, Virginia 23173} 
\affiliation{\URICH } 

\newcommand*{\SCAROLINA }{ University of South Carolina, Columbia, South Carolina 29208} 
\affiliation{\SCAROLINA } 

\newcommand*{\UNIONC }{ Union College, Schenectady, NY 12308} 
\affiliation{\UNIONC } 

\newcommand*{\VT }{ Virginia Polytechnic Institute and State University, Blacksburg, Virginia   24061-0435} 
\affiliation{\VT } 

\newcommand*{\VIRGINIA }{ University of Virginia, Charlottesville, Virginia 22901} 
\affiliation{\VIRGINIA } 

\newcommand*{\TURKEY }{ Sakarya University, Sakarya, Turkey} 
\affiliation{\TURKEY } 

\newcommand*{\YEREVAN }{ Yerevan Physics Institute, 375036 Yerevan, Armenia} 
\affiliation{\YEREVAN } 

\newcommand*{\NOWNCATU }{ North Carolina Agricultural and Technical State University, Greensboro, NC 27411}

\newcommand*{\NOWECOSSEG }{ University of Glasgow, Glasgow G12 8QQ, United Kingdom}

\newcommand*{\NOWSCAROLINA }{ University of South Carolina, Columbia, South Carolina 29208}

\newcommand*{\NOWJLAB }{ Thomas Jefferson National Accelerator Facility, Newport News, Virginia 23606}

\newcommand*{\NOWOHIOU }{ Ohio University, Athens, Ohio  45701}

\newcommand*{\NOWFIU }{ Florida International University, Miami, Florida 33199}

\newcommand*{\NOWODU }{ Old Dominion University, Norfolk, Virginia 23529}

\newcommand*{\NOWINFNFR }{ INFN, Laboratori Nazionali di Frascati, Frascati, Italy}

\newcommand*{\NOWCMU }{ Carnegie Mellon University, Pittsburgh, Pennsylvania 15213}

\newcommand*{\NOWCUA }{ Catholic University of America, Washington, D.C. 20064}

\newcommand*{\NOWINDSTRA }{ Systems Planning and Analysis, Alexandria, Virginia 22311}

\newcommand*{\NOWASU }{ Arizona State University, Tempe, Arizona 85287-1504}

\newcommand*{\NOWCISCO }{ Cisco, Washington, DC 20052}

\newcommand*{\NOWUK }{ University of Kentucky, LEXINGTON, KENTUCKY 40506}

\newcommand*{\NOWUNCW }{ North Carolina}

\newcommand*{\NOWHAMPTON }{ Hampton University, Hampton, VA 23668}

\newcommand*{\NOW }{ }

\newcommand*{\NOWTulane }{ Tulane University, New Orleans, Lousiana  70118}

\newcommand*{\NOWSACLAY }{ CEA-Saclay, Service de Physique Nucl\'eaire, F91191 Gif-sur-Yvette,Cedex, France}

\newcommand*{\NOWGEORGETOWN }{ Georgetown University, Washington, DC 20057}

\newcommand*{\NOWJMU }{ James Madison University, Harrisonburg, Virginia 22807}

\newcommand*{\NOWURICH }{ University of Richmond, Richmond, Virginia 23173}

\newcommand*{\NOWCALTECH }{ California Institute of Technology, Pasadena, California 91125}

\newcommand*{\NOWMOSCOW }{ Moscow State University, General Nuclear Physics Institute, 119899 Moscow, Russia}

\newcommand*{\NOWVIRGINIA }{ University of Virginia, Charlottesville, Virginia 22901}

\newcommand*{\NOWYEREVAN }{ Yerevan Physics Institute, 375036 Yerevan, Armenia}

\newcommand*{\NOWRICE }{ Rice University, Houston, Texas 77005-1892}

\newcommand*{\NOWINFNGE }{ INFN, Sezione di Genova, 16146 Genova, Italy}

\newcommand*{\NOWROMA }{ Universita' di ROMA III, 00146 Roma, Italy}

\newcommand*{\NOWBATES }{ MIT-Bates Linear Accelerator Center, Middleton, MA 01949}

\newcommand*{\NOWKYUNGPOOK }{ Kungpook National University, Taegu 702-701, South Korea}

\newcommand*{\NOWFSU }{ Florida State University, Tallahassee, Florida 32306}

\newcommand*{\NOWVSU }{ Virginia State University, Petersburg,Virginia 23806}

\newcommand*{\NOWORST }{ Oregon State University, Corvallis, Oregon 97331-6507}

\newcommand*{\NOWGWU }{ The George Washington University, Washington, DC 20052}

\newcommand*{\NOWMIT }{ Massachusetts Institute of Technology, Cambridge, Massachusetts  02139-4307}

\newcommand*{\NOWWM }{ College of William and Mary, Williamsburg, Virginia 23187-8795} 

%%%%%%%%%%%%%%%%%%%% authors %%%%%%%%% 
  
\author{C.~Hadjidakis}
 
     \affiliation{\ORSAY}
     \affiliation{\INFNFR}
 
\author{M.~Guidal}
 
     \affiliation{\ORSAY}
 
\author{M.~Gar\c{c}on}

     \affiliation{\SACLAY}
         
\author{J.-M. Laget}

     \affiliation{\SACLAY}
     
\author{E.S.~Smith}
 
     \affiliation{\JLAB}
     
\author{M.~Vanderhaeghen}

     \affiliation{\JLAB}
     \affiliation{\WM} 
     
\author{G.~Adams}
 
     \affiliation{\RPI}

\author{P.~Ambrozewicz}

     \affiliation{\FIU}
\author{E.~Anciant}

     \affiliation{\SACLAY}
\author{M.~Anghinolfi}

     \affiliation{\INFNGE}
% ...
f
\author{G.V.~O'Rielly}

     \affiliation{\GWU}
%\author{A.K.~Opper}

%     \affiliation{\OHIOU}
\author{M.~Osipenko}

     \affiliation{\INFNGE}
\author{K.~Park}

     \affiliation{\KYUNGPOOK}
\author{E.~Pasyuk}

     \affiliation{\ASU}
\author{G.~Peterson}

     \affiliation{\UMASS}
\author{S.A.~Philips}

     \affiliation{\GWU}
\author{N.~Pivnyuk}

     \affiliation{\ITEP}
\author{D.~Pocanic}

     \affiliation{\VIRGINIA}
\author{O.~Pogorelko}

     \affiliation{\ITEP}
\author{E.~Polli}

     \affiliation{\INFNFR}
\author{S.~Pozdniakov}

     \affiliation{\ITEP}
\author{B.M.~Preedom}

     \affiliation{\SCAROLINA}
\author{J.W.~Price}

     \affiliation{\UCLA}
\author{Y.~Prok}

     \affiliation{\VIRGINIA}
\author{D.~Protopopescu}

     \affiliation{\ECOSSEG}
\author{L.M.~Qin}

     \affiliation{\ODU}
\author{B.A.~Raue}

     \affiliation{\FIU}
     \affiliation{\JLAB}
\author{G.~Riccardi}

     \affiliation{\FSU}
\author{G.~Ricco}

     \affiliation{\INFNGE}
\author{M.~Ripani}

     \affiliation{\INFNGE}
\author{B.G.~Ritchie}

     \affiliation{\ASU}
\author{F.~Ronchetti}

     \affiliation{\INFNFR}
     \affiliation{\ROMA}
\author{P.~Rossi}

     \affiliation{\INFNFR}
\author{G.~Rosner}

     \affiliation{\ECOSSEG}
\author{D.~Rowntree}

     \affiliation{\MIT}
\author{P.D.~Rubin}

     \affiliation{\URICH}
\author{F.~Sabati\'e}

     \affiliation{\SACLAY}
     \affiliation{\ODU}
\author{K.~Sabourov}

     \affiliation{\DUKE}
\author{C.~Salgado}

     \affiliation{\NSU}
\author{J.P.~Santoro}

     \affiliation{\VT}
     \affiliation{\JLAB}
\author{V.~Sapunenko}

     \affiliation{\INFNGE}
\author{R.A.~Schumacher}

     \affiliation{\CMU}
\author{V.S.~Serov}

     \affiliation{\ITEP}
\author{A.~Shafi}

     \affiliation{\GWU}
\author{Y.G.~Sharabian}
 
%      \altaffiliation[Current address:]{\NOWJLAB}

     \affiliation{\JLAB}
     \affiliation{\YEREVAN}
\author{J.~Shaw}

     \affiliation{\UMASS}
\author{S.~Simionatto}

     \affiliation{\GWU}
\author{A.V.~Skabelin}

     \affiliation{\MIT}

\author{L.C.~Smith}

     \affiliation{\VIRGINIA}
\author{D.I.~Sober}

     \affiliation{\CUA}
\author{M.~Spraker}

     \affiliation{\DUKE}
\author{A.~Stavinsky}

     \affiliation{\ITEP}
\author{S.~Stepanyan}
 
%      \altaffiliation[Current address:]{\NOWODU}

	\affiliation{\JLAB}
        \affiliation{\YEREVAN}

\author{B.E.~Stokes}

     \affiliation{\FSU}
\author{P.~Stoler}

     \affiliation{\RPI}
%\author{I.I.~Strakovsky}

%     \affiliation{\GWU}
\author{S.~Strauch}

     \affiliation{\GWU}
\author{M.~Taiuti}

     \affiliation{\INFNGE}
\author{S.~Taylor}

     \affiliation{\RICE}
\author{D.J.~Tedeschi}

     \affiliation{\SCAROLINA}
\author{U.~Thoma}

     \affiliation{\GEISSEN}
     \affiliation{\JLAB}
\author{R.~Thompson}

     \affiliation{\PITT}
\author{A.~Tkabladze}

     \affiliation{\OHIOU}
\author{L.~Todor}

     \affiliation{\URICH}
\author{C.~Tur}

     \affiliation{\SCAROLINA}
\author{M.~Ungaro}

     \affiliation{\RPI}
\author{M.F.~Vineyard}

     \affiliation{\UNIONC}
     \affiliation{\URICH}
\author{A.V.~Vlassov}

     \affiliation{\ITEP}
\author{K.~Wang}

     \affiliation{\VIRGINIA}
\author{L.B.~Weinstein}

     \affiliation{\ODU}
%\author{A.~Weisberg}

%     \affiliation{\OHIOU}
\author{H.~Weller}

     \affiliation{\DUKE}
\author{D.P.~Weygand}

     \affiliation{\JLAB}
\author{C.S.~Whisnant}
 
%      \altaffiliation[Current address:]{\NOWJMU}
      
     \affiliation{\JMU}
     \affiliation{\SCAROLINA}
\author{M.~Williams}

     \affiliation{\CMU}
\author{E.~Wolin}

     \affiliation{\JLAB}
\author{M.H.~Wood}

     \affiliation{\SCAROLINA}
\author{A.~Yegneswaran}

     \affiliation{\JLAB}
\author{J.~Yun}

     \affiliation{\ODU}
\author{L.~Zana}

     \affiliation{\UNH}

\collaboration{The CLAS Collaboration}
     \noaffiliation
% 
 
%The Southeastern Universities Research Association (SURA) operates the 
%Thomas Jefferson National Accelerator Facility for the United States 
%Department of Energy under contract DE-AC05-84ER40150. 
% 
% 
%%%%%%%%%%%%%%%%%%%%%%%%%%%%%%%%%%% 
% FOLLOWING A GENERIC TAIL FOR TEX
%   Nothing here relates to author list
%   Suggestion for acknowledgements.  Could be deleted.
\date{\today}
\begin{abstract}
The longitudinal and transverse components of the cross section for
the $e p\to e^\prime p \rho^0$ reaction were measured in Hall B
at Jefferson Laboratory using the CLAS 
detector. The data were taken with a 4.247 GeV electron beam 
and were analyzed in a range of $x_B$ from 0.2 to 0.6 and of $Q^2$
from 1.5 to 3.0 GeV$^2$. The data are compared to a Regge model
based on effective hadronic degrees of freedom and to a 
calculation based on Generalized Parton Distributions.
It is found that the transverse part of the cross section is well
described by the former approach while the longitudinal part can be
reproduced by the latter. 
\end{abstract}
\maketitle

\begin{figure}[h]
%\vspace{-3.cm}
%\includegraphics[width=8.5cm,height=12cm]{diags.ps}
\includegraphics[width=1.\columnwidth,bb=75 375 550 600,clip=true]{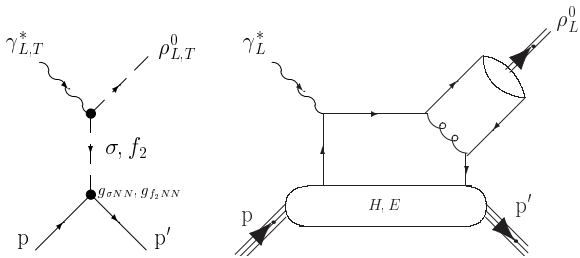}
%\vspace{-3.25cm}
\caption[]{The mechanisms for $\rho^0$ electroproduction at intermediate 
energies: at low $Q^2$ (left diagram) through the exchange of mesons, and 
at high $Q^2$ (right diagram) through the quark exchange ``handbag" mechanism
(valid for longitudinal photons) where $H$ and $E$ are the unpolarized GPD's.}
\label{fig:diags}
\end{figure}

Understanding the precise nature of the confinement of 
quarks and gluons inside hadrons has been an 
%ever-lasting 
ongoing problem
since the advent, about 30 years ago, of the theory that governs their 
interactions, quantum chromodynamics (QCD). In particular, the
transition between the high energy (small distance) domain,
where quarks are quasi-free, and the low energy (large distance) 
regime, where they form bound states and are confined in hadrons, 
is still not well understood.
\newline
\indent 
The analysis of elementary processes, such as the exclusive electroproduction 
of a meson or a photon on the nucleon in the few GeV range, allows one to 
study this transition. In the case of exclusive meson electroproduction,
the longitudinal and transverse polarizations of the (virtual)
photon mediating the interaction provide two
qualitatively different pieces of information about the nucleon structure.
\newline
\indent 
Longitudinal photons, whose transverse size is inversely
proportional to their virtuality, truly act as a microscope. 
At sufficiently large $Q^2$, small distances are probed, and
the asymptotic freedom of QCD justifies the understanding
of the process in terms of partonic degrees of freedom and the use of 
perturbative QCD (pQCD) techniques. In particular, it has been recently
shown~\cite{Collins97,frank97} that the non-perturbative information can be 
factorized in reactions such as exclusive 
vector meson electroproduction. Here the process can be described in 
terms of perturbative quark or gluon exchanges whose momentum, flavor, 
and spin distributions inside the nucleon are parametrized in terms of the 
recently introduced Generalized Parton Distributions
(GPD's)~\cite{Ji97,Rady,Mul94}. This is the so-called 
``handbag" diagram mechanism which is depicted in
Fig.~\ref{fig:diags} (right diagram). At higher $\gamma^*p$ center-of-mass 
energies, $W$, than considered in this letter, 2-gluon exchange processes
also intervene~\cite{frank97,param_GPD}.
At low virtuality, $Q^2$, of the photon, hadronic degrees of freedom
are more relevant and, above the nucleon resonance region, the process 
is adequately described in terms of meson exchanges (Fig.~\ref{fig:diags}, 
left diagram).
\newline
\indent 
For transverse photons, however, this description in terms of 
quarks and gluons is not valid. A factorization into a hard and 
soft part does not hold~\cite{Collins97,frank97} and
even at large $Q^2$, there is no dominance 
of a ``handbag" mechanism as in Fig.~\ref{fig:diags}. ``Soft" 
(non-perturbative) and ``hard" (perturbative) physics compete over 
a wider range of $Q^2$, and in practice it is necessary to take into account 
non-perturbative effects using hadron degrees of freedom.
In order to access the fundamental partonic information when studying 
meson electroproduction processes, it is therefore highly 
desirable to isolate the 
longitudinal part of the cross section, which lends itself, at least at
sufficiently high $Q^2$, to pQCD techniques and interpretation.
In this approach, however, several questions remain to be answered. 
What is the lowest $Q^2$ where a perturbative
treatment is valid? What corrections need to be applied to
extend its validity to lower $Q^2$ ?

\begin{figure}[h]
%\vspace{-1.cm}
\includegraphics[width=1.\columnwidth,clip=true]{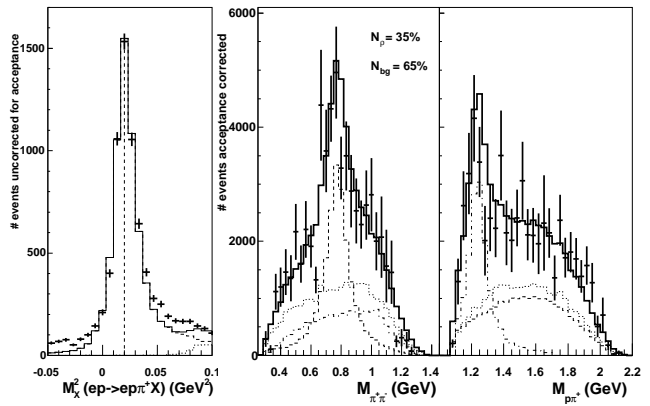}
%\vspace{-0.5cm}
\caption[]{Left plot: an example of a squared missing mass 
$M^2_X(ep\to e^\prime p\pi^+X)$ 
spectrum (for a scattered electron momentum between 1.9 and 2.2 GeV). 
Points with error bars show the experimental data and the solid lines represent 
the results of simulations for the channels $e^\prime p \pi^+\pi^-$ 
(dashed line), $e^\prime p \pi^+\pi^-\pi^0$ (dotted line)
and the sum of the two (solid line). The vertical dashed line
is located at the missing mass squared of a pion. Central and right plots: 
an example of the $\pi^+\pi^-$ and $p\pi^+$ invariant masses, 
respectively (for the interval 1.63 $< Q^2 <$ 1.76 GeV$^2$ and 
0.28 $< x_B <$ 0.35). Points with error bars show the experimental data and
the lines correspond to the results of fits for the channels 
$e p\to e^\prime p \rho^0$ (dashed line), 
$e p\to e^\prime \Delta^{++} \pi^-$ (dash-dotted line), 
non-resonant $e p\to e^\prime p \pi^+ \pi^-$ (dotted line) and
the sum of the three processes (solid line).}
\label{fig:im}
\end{figure}

The aim of this letter is to address these
questions using the recent measurement of the longitudinal
and transverse cross sections of the $e p\to e^\prime p \rho^0$ reaction,
carried out at Jefferson Laboratory 
using the CEBAF Large Acceptance Spectrometer (CLAS)~\cite{clas} in Hall B.
This elementary process is one of the exclusive reactions on the nucleon 
which has the highest cross section, and for which the extraction of the 
longitudinal and transverse parts of the cross section 
can be accomplished using the $\rho^0$ decay angular distribution. 
On the theoretical side, formalisms and numerical estimates for both 
hadronic and partonic descriptions of the reaction have been developed, 
which can be compared to the transverse and longitudinal components
of the cross section, respectively.
\newline
\indent
In the following, we will present the analysis results of the 
$e p\to e^\prime p \rho^0$ reaction. Data were taken with an 
electron beam energy of 4.247 GeV impinging on an unpolarized liquid-hydrogen
target. 
The integrated luminosity of this data set was about 1.5 fb$^{-1}$.
The kinematic domain of the selected sample corresponds to 
$Q^2$ from 1.5 GeV$^2$ to 3.0 GeV$^2$. We analyzed data for
$W$ greater than 1.75 GeV, which 
corresponds to a range of $x_B$ from 0.21 to 0.62.
Our final data sample included about $2\times 10^4$ 
$e^\prime p \pi^+ \pi^-$ events. 
\newline
\indent
The $\rho^0$ meson decay to $\pi^+\pi^-$ was used to identify
the reaction of interest. We identified the 
$e p\to e^\prime p \pi^+ \pi^-$ reaction using the missing
mass technique by detecting the scattered electron, the recoil proton, and 
the positive pion. The electron was identified as a negative track with 
reconstructed energy deposition in the calorimeter which was consistent with the
momenta determined from magnetic analysis, in combination with a signal
in the Cerenkov counter. The proton and pion were identified as positive tracks,
whose combination of flight times and momenta corresponded to their mass.
Figure\,\ref{fig:im} (left plot) shows a typical 
missing mass distribution for $e p\to e^\prime p \pi^+ X$ events. 
Events were selected by the missing mass
cut -0.03 $< M^2_X <$ 0.06 GeV$^2$, consistent with a missing $\pi^-$. 
Fig.\,\ref{fig:im} (center) shows the
resulting $\pi^+\pi^-$ invariant mass spectrum. The $\rho^0$ peak is 
clearly visible, sitting on a large non-resonant 
$\pi^+\pi^-$ background. 
\newline
\indent

\begin{figure}[h]
\vspace{-.4cm}
\includegraphics[width=1.\columnwidth,clip=true]{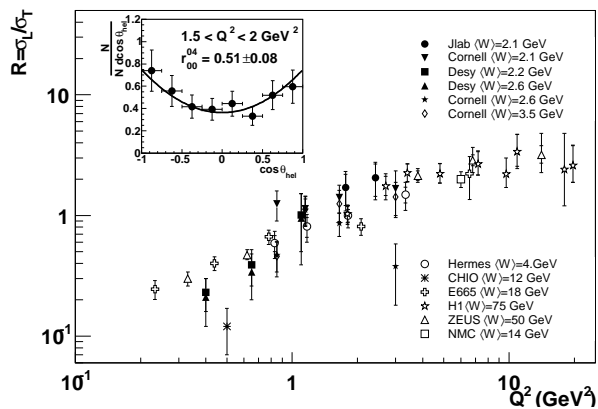}
\vspace{-.4cm}
\caption[]{
%The ratio R=$\sigma_L$/$\sigma_T$ as a function of $Q^2$ 
%for $W\approx$ 2.1 GeV. The other data are from DESY~\cite{desy}
%and Cornell~\cite{cornell}. 
The ratio R= $\sigma_L$/$\sigma_T$ as a function of $Q^2$ for $\rho^0$ meson 
electroproduction on the nucleon. The other data are from ~\cite{desy,cornell,
hermes2,hermes3,chio,fermi,h1,zeus,nmc}. The  insert shows one of our 
$\cos\theta_{\rm hel}$ distributions with a fit to determine $r^{04}_{00}$.}
\label{fig:r04}
\end{figure}

The unpolarized 
$e p\to e^\prime p \pi^+ \pi^-$ reaction is fully defined by seven independent
kinematical variables which we have chosen as: $Q^{2}$ and $x_{B}$, which
define the virtual photon kinematics; $t$, the invariant squared momentum transfer
between the virtual photon and the final pion pair ($i.e.$ the $\rho^0$ meson when
this particle is produced); $M_{\pi^+\pi^-}$, the invariant
mass of the $\pi^+\pi^-$ system; $\theta_{\rm hel}$ and $\phi_{\rm hel}$, 
the $\pi^+$ decay angles in the $\pi^+\pi^-$ rest frame; and $\Phi$, the
azimuthal angle between the hadronic and leptonic planes.
The CLAS acceptance and efficiency were calculated for
each of these 7-dimensional bins using a 
GEANT-based simulation of several hundred million events. The event distributions
were generated according to Ref.~\cite{genovamc}, which includes the 
three main contributions above the resonance region to the
$e^\prime p \pi^+ \pi^-$ final state: diffractive $e p\to e^\prime p \rho^0$, 
$t$-channel $e p\to e^\prime \Delta^{++}\pi^-$, and 
non-resonant (phase space) $e p\to e^\prime p \pi^+ \pi^-$. Each of these contributions 
to the event generator was  
matched to the world's data on differential and total cross sections,
and then extrapolated to our kinematical domain. We were then able to extract 
a total cross section for the $e p\to e^\prime p \pi^+ \pi^-$ channel in good agreement
with world's data where the kinematics overlapped.
The event generator also includes radiative effects following the Mo and 
Tsai prescription~\cite{MoandTsai} so that radiative corrections could 
be applied in each ($Q^2$, $x_B$) bin. 
\newline
\indent
The main difficulty in determining the $\rho^0$ yield stems from
its large width ($\Gamma_{\rho^0}\sim 150$ MeV),
which does not allow for a unique determination of 
the separate contributions due to
the resonant $\rho^0$ production and 
non-resonant $\pi^+\pi^-$ pairs. We simultaneously fitted the two 3-fold
differential cross sections $d^3\sigma/dQ^2dx_BdM_{\pi^+\pi^-}$ and
$d^3\sigma/dQ^2dx_BdM_{p\pi^+}$ to determine the weight of the three channels
mentioned earlier, leading to the $e^\prime p \pi^+ \pi^-$ final state 
(see Fig.~\ref{fig:im}, central and right plots). The mass spectra of the 
$\rho^0$ and $\Delta^{++}$ are generated according to standard Breit-Wigner
distributions
and the non-resonant $p \pi^+ \pi^-$ final state according to phase space.
This background estimation procedure, along with the CLAS acceptance modeling, 
is one of the dominant sources of systematic
uncertainty which, in total, ranges from 10\% to 25\%. More sophisticated 
shapes for the $\rho^0$ mass spectra were also investigated
but led to consistent numbers of $\rho^0$'s within these error bars.
\newline
\indent
The final step of the analysis consisted in separating the longitudinal
and the transverse parts of the $e p\to e^\prime p \rho^0$ cross section.
The determination of these two contributions was
accomplished under the assumption of $s$-channel helicity conservation (SCHC) 
\cite{Bauer}. This hypothesis states, in simple terms, that the helicity 
of the 
%vector meson is directly transferred to the virtual photon. 
virtual photon is directly transferred to the vector meson.	
The SCHC hypothesis originates
from the vector meson dominance model which identifies vector meson
electromagnetic production as an elastic process without spin transfer.
\newline
\indent
The validity of the SCHC hypothesis, which is only applicable
at small momentum transfer $t$, can be tested experimentally through the analysis 
of the azimuthal angular distribution. We found that the $r^{04}_{1-1}$ 
$\rho^0$ decay matrix element~\cite{schilling}, which can be extracted 
from the $\phi_{\rm hel}$ dependence, was compatible with zero at the 
1.7 sigma level. We also found that the $\sigma_{TT}$ and $\sigma_{TL}$ 
cross sections, which can be extracted from the $\Phi$ dependence,
were, respectively, 10.6\% $\pm$ 11.8\% and 0.4\% $\pm$ 5.4\% of
the total cross section. They are therefore consistent with zero, as they 
should be if SCHC is valid and, in any case, don't represent potential
large violations of SCHC. Let's also note that all previous experiments 
on electromagnetic production of $\rho^0$ on the nucleon 
are consistent with the dominance of $s$-channel helicity conserving amplitudes
(the helicity-flip amplitudes which have been reported~\cite{desy,h1,zeus,hermes2}
never exceeded 10-20\% of the helicity non-flip amplitudes). We can
therefore safely rely on SCHC for our analysis.
\newline
\indent
The decay angular distribution of the $\pi^+$
in the $\rho^0$ rest frame can be written as~\cite{schilling}: 
\begin{eqnarray}
W(\cos\theta_{\rm hel})=\frac{3}{4}\left[ 1-r^{04}_{00}+(3r^{04}_{00}-1)\cos^2\theta_{\rm hel} 
\right],
\label{eq:r04_rho}
\end{eqnarray}
where $r^{04}_{00}$ represents the degree of longitudinal polarization of 
the $\rho$ meson. Under the assumption of SCHC,
the ratio of longitudinal to transverse cross sections is:
\begin{eqnarray}
R_\rho=\frac{\sigma_L}{\sigma_T}=\frac{1}{\epsilon}\frac{r^{04}_{00}}
{1-r^{04}_{00}},
\label{eq:r_rho}
\end{eqnarray}
where $\epsilon$ is the virtual photon transverse 
polarization. 
$r^{04}_{00}$ was extracted from the fit of the background-subtracted 
$\cos\theta_{\rm hel}$ distributions
following Eq.~\ref{eq:r04_rho} 
as illustrated in the insert in Fig.~\ref{fig:r04}, 
and was used in Eq.~\ref{eq:r_rho} to determine
$R_\rho$.

%Due to limited statistics, this procedure could be performed only for the 
%two $Q^2$ points which are shown on Fig.~\ref{fig:r04} with
%the existing world data in a 
%similar $W$ range ($W\approx$ 2.1 GeV)~\cite{desy,cornell},
%which also entered our fit.  
%We then were able to obtain the following parametrization of $R_\rho$ as a 
%function of $Q^2$: $R_\rho$=0.75$\times (Q^2)^{1.1}$.
%We note that by comparisons with the world's data~\cite{hermes,fermi}
%at higher energy, this parametrization changes with $W$.

Due to limited statistics in the CLAS data, this procedure could 
be performed only for the 
two $Q^2$ points which are shown on Fig.~\ref{fig:r04} and where our points
are found to be compatible with the existing world's data. We then have
fitted the $Q^2$ dependence of $R_\rho$ including, in 
order to take into account a potential $W$ dependence of the ratio $R$, 
only the world's data in the $W$ domain close to ours
($W\approx$ 2.1 GeV)~\cite{desy,cornell}. The
following parametrization, whose power form is motivated by
the perturbative PQCD prediction that $\sigma_T$ is power suppressed
with respect to $\sigma_L$, was found : 
\begin{eqnarray}
R_\rho=0.75\pm 0.08 \times (Q^2)^{1.09\pm 0.14}.
\end{eqnarray}

\begin{figure}[h]
\includegraphics[width=1.\columnwidth,clip=true]{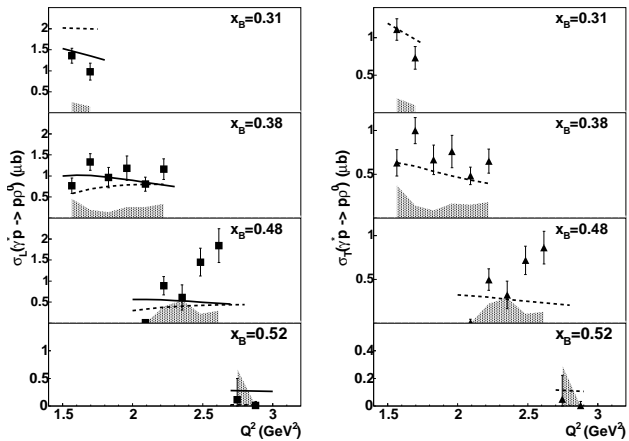}
\vspace{-0.4cm}
\caption[]{Cross sections $\sigma_L$ (left) and $\sigma_T$ (right) 
for $e p\to e^\prime p \rho^0$ as a function of $Q^2$ as 
measured in this experiment. The dotted line represents the
Regge model of Refs.~\cite{laget1,laget2}
while the solid line describes the GPD model of Refs.~\cite{param_GPD,Goeke}.
The systematic error is indicated by the shaded zones
at the bottom of the plots.}
\label{fig:siglt}
\end{figure}

It is customary to define the
reduced cross section for $\rho$ meson production as the
electroproduction cross section divided by the flux of virtual photons~:
\begin{eqnarray}
\sigma_T +\epsilon \sigma_L=\frac{1}{\Gamma_V(Q^2,x_B)}\times\frac{d^2\sigma^{e p}}{dQ^2dx_B},
\label{eq:sigtot}
\end{eqnarray}
where the virtual photon flux is given by~:
\begin{eqnarray}
\Gamma_V(Q^2,x_B)=\frac{\alpha}{8\pi}\frac{Q^2}{M_p^2E_e^2}
\frac{1-x_B}{x_B^3}\frac{1}{1-\epsilon}.
\label{eq:sigtot2}
\end{eqnarray}
In this notation, and in Fig.~\ref{fig:siglt}, the longitudinal and transverse
$\sigma_T$ and $\sigma_L$ cross sections are
integrated over $t$, $\Phi$, $\theta_{\rm hel}$, and $\phi_{\rm hel}$. The
$t$ dependence of $\sigma_T +\epsilon \sigma_L$ can be parametrized
by $e^{-b | t-t_{\rm min}|}$ ($-t_{\rm min} < -t < 1$ GeV$^2$),
where $-t_{\rm min}$ is the smallest value of momentum transfer for a given kinematic bin.
We measured the exponential slope $b$ to range from 1.19 to 1.74  GeV$^{-2}$
for $x_B$ between 0.31 and 0.52.

The longitudinal and transverse cross sections are plotted in Fig.\,\ref{fig:siglt} 
as a function of $Q^2$ for four bins centered 
at $x_B$ of 0.31, 0.38, 0.48, and 0.52. These values correspond to $W$ values of 2.2, 
2.0, 1.9, and 1.85 GeV, respectively. 
The data are compared to two theoretical approaches. The first
one is based on hadronic degrees of freedom with meson 
Regge trajectory exchanges in the $t$-channel (as illustrated
in Fig.~\ref{fig:diags}, left graph). 
This approach has been successful in describing, with very few free parameters,
essentially all of the available observables of a series of forward exclusive reactions 
in photo- and electroproduction of pseudoscalar mesons ($\pi^{0,\pm}$, 
$K^+$~\cite{ourpapers}, $\eta, \eta^\prime$~\cite{taiwan}) above the resonance 
region. For the $\rho^0, \omega, \phi$ vector mesons, as well as for Compton 
scattering, such an approach has been recently developed in Refs.~\cite{laget1,
laget2,laget3}. 
In the case of $\rho^0$ electroproduction, the contributing meson trajectories are 
the $\sigma$, $f_2$, and Pomeron, the latter being negligible in the $W$ region
investigated in this experiment. This Regge model was normalized by adjusting
the $\sigma$ and $f_2$ meson-nucleon couplings to reproduce
existing photoproduction data (see for instance, Refs.~\cite{batta}).
There is little freedom in the choice of parameters when one uses data
from all three $\rho^0, \omega$, and $\phi$ channels, which together constrain 
all photoproduction parameters. The only remaining free parameters
for the electroproduction case are 
the squared mass scales of the meson monopole form factors at the electromagnetic 
vertices for the diagrams of Fig.~\ref{fig:diags} (left plot). They have been 
determined from the $Q^2$ dependence of the world's data, in
particular from the Cornell~\cite{cornell} and HERMES~\cite{hermes} experiments,
to be approximatively 0.5 GeV$^2$, in accordance with known meson form 
factor mass scales.
\newline
\indent
As shown in Fig.~\ref{fig:siglt}, this Regge model provides 
a fair description of the transverse and longitudinal cross sections. 
There is some discrepancy at large values of $x_B$, but at those values 
some $s$-channel nucleonic resonances decaying into $\rho^0 p$ may contribute, a process
which is not taken into account in this Regge $t$-channel approach, and might explain 
the missing strength in this particular kinematical domain.
The calculation was also done for the Cornell~\cite{cornell} and
HERMES~\cite{hermes} data, where, general agreement is found as well 
(the longitudinal cross section is also overestimated as one goes 
to smaller $x_B$, as in our $x_B$= 0.31 bin).
\newline
\indent
We now turn to the handbag diagram approach (Fig.~\ref{fig:diags}, right plot), 
which is based on the QCD factorization between a ``hard" process (the interaction between a quark 
of the nucleon and the virtual photon, along with a one-gluon exchange for the formation
of the final meson) and a ``soft" process (the parametrization of the partonic 
structure of the nucleon in terms of GPD's). As mentioned in the introduction, 
this approach is only valid 
at sufficiently large $Q^2$ when the longitudinal cross section dominates
the QCD expansion in powers of $1/Q^2$. Unfortunately, the value of $Q^2$ at which the
``handbag" mechanism becomes valid is unknown, and especially for meson electroproduction,
it must be determined experimentally.
\newline
\indent
In the case of $\rho^0$ production, only the unpolarized GPD's $H$ and 
$E$ contribute to the amplitude of the reaction. In the calculation, 
shown in Fig.~\ref{fig:siglt}, we neglect the contribution due to
the GPD $E$ because it is proportional to the 4-momentum transfer between 
the incoming virtual photon and the outgoing meson, and our data cover small
momentum transfers. For the GPD $H$ we use 
the parametrization of Refs.~\cite{param_GPD,Goeke}. The other ingredient 
entering the (leading order) calculation of the handbag diagram is the treatment of 
the strong coupling constant $\alpha_s$ between the quarks and the gluon.
It has been ``frozen" to a value of 0.56, as determined
by QCD sum rules~\cite{Bal96}. The freezing
of the strong coupling constant $\alpha_s$ is an effective way to
average out non-perturbative effects at low $Q^2$ and is supported by jet-shape
analysis of the infrared coupling~\cite{dok}.
\newline
\indent
As mentioned earlier, the handbag diagram calculation can only be compared
with the longitudinal part of the cross section.
Figure~\ref{fig:siglt} shows a good agreement between 
the calculation and the data at the low $x_B$ values. As for the Regge model
discussed above, the two highest $x_B$ bins might contain some additional
nucleonic resonance ``contamination", which are not included in the ``handbag"
approach. Variations in 
reasonable ranges of the parameters entering the GPD's were studied, and results
were found to be stable at the 50\% level. This provides confidence in 
the stability, reliability, and validity of the calculation based on the prescription of a
``frozen" $\alpha_s$. Let us also note that this calculation
reproduces reasonably well the HERMES data~\cite{hermes}, which were taken at 
neighboring kinematics.
\newline
\indent 
A signature of the handbag mechanism is that, independent of any
particular GPD parametrization adopted, the (reduced)
cross sections should follow a $1/Q^6$ dependence at fixed $t$ and $x_B$. 
In this analysis, due to the lack of statistics, $\sigma_L$ is integrated over $t$, which  
means that it is proportional to $t_{\rm min}$, this latter variable 
changing as a function of $Q^2$. 
%\footnote{The determination of the dependence of the cross sections as a function of
%$Q^2$ at fixed values of $t$ and fixed $x_B$, which is one of the ultimate goal
%of this program, is impractical due to the lack of statistics in the present experiment.}. 
This $1/Q^6$ scaling behavior at fixed $t$ and $x_B$ can therefore not
be directly observed in our data, which is modified by the
(trivial) kinematical $Q^2$ dependence of $t_{\rm min}$. Nevertheless, agreement between
the data and the GPD calculation, which also contains this trivial $t_{\rm min}$ dependence,
should be interpreted as confirmation of the  leading order prediction based on the
``handbag'' diagram.
\newline
\indent 
In conclusion, we have presented here a first exploration of exclusive
vector meson electroproduction on the nucleon 
in a region of $Q^2$ between 1.5 and 3.0 GeV$^2$ and $x_{B}$ between 
0.2 and 0.6, which is a kinematical domain barely explored.
The Regge model, based on ``economical'' hadronic degrees of freedom, is able to describe
% this new notation, should not be introduced in the conclusion. 
%using a few effective 
%parameters such as the $g_{\sigma NN}*g_{\sigma \rho \gamma}$ and $f_2 NN$ coupling constants
%and meson form factor mass scales, 
the transverse cross section data, along with the
other existing vector meson photo- and electroproduction data.
Furthermore, the more fundamental ``handbag'' approach, with a standard parametrization
of the GPD $H$ and the extrapolation to low $Q^2$ by an effective freezing of $\alpha_s$, 
provides a fair description of the longitudinal part of 
the cross section. Therefore it seems possible to understand the longitudinal
part of the $\rho$ meson production cross section in a pQCD framework, which potentially 
gives access to GPD's. The transverse cross section, on the other hand, for which no factorization
between soft and hard physics exists, can be described
in terms of meson exchanges.
These tentative conclusions need of course to be confirmed by a more extensive 
and thorough exploration of the $x_B,Q^2$ phase space which is currently 
under way with a much larger data set~\cite{propo}.

We would like to acknowledge the outstanding efforts of the staff of the 
Accelerator and the Physics Divisions at Jefferson Lab that made this experiment possible.
This work was supported in part by the Istituto Nazionale di Fisica Nucleare, the 
 French Centre National de la Recherche Scientifique, 
the French Commissariat \`{a} l'Energie Atomique, the U.S. Department of Energy, the National 
Science Foundation, Emmy Noether grant from the Deutsche Forschungsgemeinschaft and the Korean Science and Engineering Foundation.
The Southeastern Universities Research Association (SURA) operates the 
Thomas Jefferson National Accelerator Facility for the United States 
Department of Energy under contract DE-AC05-84ER40150.

\end{document}